\documentclass[11pt]{article}
\usepackage{amssymb}
\usepackage{graphics}
\usepackage{epsfig}
\usepackage{a4wide}
\usepackage{slashed}
\usepackage{multicol,bbm}
\usepackage{multirow}

\begin{document}

\title{Probing $L$-violating coupling via sbottom resonance production at the LHeC}
\author{ Zhang Ren-You, Wei Hua, Han Liang and Ma Wen-Gan \\
{\small  Department of Modern Physics, University of Science and Technology of China (USTC),}  \\
{\small  Hefei, Anhui 230026, People's Republic of China}}

\date{}
\maketitle \vskip 8mm
\begin{abstract}
We investigate the resonant production of the lighter sbottom at the proposed Large Hadron Electron Collider
and its subsequent decay into $e^- + jet$ final state, under the single coupling dominance hypothesis in the framework
of the $R$-parity-violating minimal supersymmetric standard model. It is concluded that the LHeC running
at the high electron beam energy configuration of $E_e = 150~{\rm GeV}$ would provide excellent opportunities to search
for the sbottom resonance and probe the lepton-number-violating $\hat{L}\hat{Q}\hat{D}$ $\lambda^{\prime}_{113}$-type
Yukawa interaction. With about $1~fb^{-1}$ integrated luminosity expected in $E_e = 150~{\rm GeV}$ $e^-p$ collision at
the LHeC, either the lighter sbottom RPV resonance could be directly detected up to $1~{\rm TeV}$, or the $L$-violating
coupling $\lambda^{\prime}_{113}$ could be constrained at an unprecedented level of precision compared with all the
knowledge derived from indirect measurements.
\end{abstract}

\vskip 8mm
{\large\bf PACS: 12.60.Jv, 14.80.Ly, 11.30.Fs, 13.85.Rm}

\vfill \eject \baselineskip=0.32in

\renewcommand{\theequation}{\arabic{section}.\arabic{equation}}
\renewcommand{\thesection}{\Roman{section}.}
\newcommand{\nb}{\nonumber}

\newcommand{\Dir}{\kern -6.4pt\Big{/}}
\newcommand{\Dirin}{\kern -10.4pt\Big{/}\kern 4.4pt}
\newcommand{\DDir}{\kern -7.6pt\Big{/}}
\newcommand{\DGir}{\kern -6.0pt\Big{/}}

\makeatletter      
\@addtoreset{equation}{section}
\makeatother       

\section{Introduction}
\par
The standard model (SM) \cite{SM-1,SM-2} provides a remarkably successful description of strong, weak and electromagnetic
interactions at the energy scale up to $10^2~{\rm GeV}$. It implies the conservation of baryon number $B$ and lepton number
$L$ as the automatic consequence of gauge invariance and renormalizability. Despite its tremendous success, the SM suffers
from some conceptional difficulties, such as the hierarchy problem and the nonoccurrence of gauge coupling unification at
high energy scale. Supersymmetry theories may provide solutions for such problems, and are widely expected as the most
appealing extensions of the SM. The simplest supersymmetric extension of the SM with minimal particle content is called
minimal supersymmetric standard model (MSSM) \cite{Nilles,Haber,Gunion}. To prevent proton from rapid decay \cite{Sakai,Dimopoulos},
a multiplicative $R$-parity is introduced as $R_p = (-1)^{3B+L+2S}$ in terms of $B$, $L$ and spin quantum number $S$ \cite{Fayet-1,Fayet-2,Fayet-3},
which leaves all the SM particles $R_p = +1$ while for their superpartners $R_p = -1$. The conservation of the $R$-parity
imposes strict simultaneous $B$ and $L$ symmetry, and results in a stable lightest supersymmetric particle (LSP), which can
be a candidate for dark matter if it is color and electrically neutral. However, such a stringent symmetry appears to be
short of theoretical basis, especially as it is known that a stable proton can survive by requiring only either $B$ or $L$
conservation and partial lepton-flavor violation should be introduced to accommodate neutrino oscillations.
Consequently, the $R$-parity-violating (RPV) interactions could be included into the most general representation
of the superpotential as \cite{Weinberg}
\begin{eqnarray}
  W_{\slashed{R}_{p}} =
  \frac{1}{2} \lambda_{ijk} \hat{L}_i\hat{L}_j \hat{E}^c_k
  + \lambda^{\prime}_{ijk} \hat{L}_i \hat{Q}_j \hat{D}^c_k
  + \frac{1}{2} \lambda^{\prime \prime}_{ijk} \hat{U}^c_i \hat{D}^c_j \hat{D}^c_k
  + \mu_i \hat{H}_u \hat{L}_i,
\end{eqnarray}
where $i,j,k = (1,2,3)$ are generation indices, while $SU(2)$ isospin and $SU(3)$ color indices are suppressed. $\hat{L}$
and $\hat{Q}$ are the lepton and quark $SU(2)$ left-handed doublet chiral superfields, $\hat{E}$, $\hat{U}$ and $\hat{D}$
are the right-handed singlets, and $\hat{H}_{u}$ is the left-handed up-type Higgs doublet. The bilinear terms $\hat{H}\hat{L}$,
denoting the mixture of Higgs and lepton superfields, will generate neutrino masses in a natural way. The trilinear terms
$\hat{L}\hat{L}\hat{E}$ and $\hat{L}\hat{Q}\hat{D}$ with dimensionless RPV Yukawa couplings $\lambda$ and $\lambda^{\prime}$
only violate $L$ symmetry, while the $\lambda^{\prime \prime}$-parameterized trilinear terms $\hat{U}\hat{D}\hat{D}$ only
break baryon-number conservation. The constraints on the RPV couplings have been derived by precise measurements of low
energy processes and decays, and are well summarized in the literature \cite{Barbier}.

\par
The most distinctive phenomenology of the $R$-parity-violating MSSM ($\slashed{R}_{p}$-MSSM) is that, the LSP is no longer
stable and is not necessarily an electrically neutral and uncolored particle. As a consequence, supersymmetric scalar leptons
and quarks, such as the lighter sbottom $\tilde{b}_{1}$, can be candidates for the LSP, and the massive scalar fermion LSP
can only decays into ordinary SM particles after its single/pair production at high erergy colliders. As we know, the bottom squark
could be pair-produced at the Large Hadron Collider (LHC) via QCD interactions. The production rates and the differential
distributions with respect to some kinematical variables, including the next-to-leading order QCD corrections \cite{Beenakker} and the electroweak
contributions \cite{Germer}, have been calculated in several SUSY scenarios.
As shown in ${\rm Table~ \ref{table0-LOcs-LHC}}$, the cross section for the sbottom pair production at the LHC decreases
by $5 - 7$ orders of magnitude as the increment of $m_{\tilde{b}}$ from $100$ to about $1000~ {\rm GeV}$, due to the drastic
reduction of the final state phase space and the smallness of parton densities in proton at large $x$.
\begin{table}[htb]
  \centering
  \begin{tabular}{|c|c|c|}
  \hline
  \multirow{2}*{$m_{\tilde{b}}~({\rm GeV})$} & \multicolumn{2}{|c|}{$\sigma(pp \rightarrow \tilde{b} \tilde{b}^{*} + X)~ (pb)$} \\
  \cline{2-3}
   & ~~~~LHC @ 7 TeV~~~~ & ~~~~LHC @ 14 TeV~~~~ \\
  \hline
  100 & 305 & $1.35 \times 10^3$ \\
  400 & 0.156 & 1.67 \\
  500 & 0.03042(2) & 0.4443(3) \\
  633 & $5.175(2) \times 10^{-3}$ & 0.11236(6) \\
  1070 & $3.706(1) \times 10^{-5}$ & $3.405(1) \times 10^{-3}$ \\
  \hline
  \end{tabular}
  \caption{\label{table0-LOcs-LHC}
  Leading order cross sections for sbottom-antisbottom production at the LHC. The results for $m_{\tilde{b}} = 100,~ 400~{\rm GeV}$ and
  $m_{\tilde{b}} = 500,~ 633,~ 1070~{\rm GeV}$ are quoted from Refs.\cite{Beenakker} and \cite{Germer}, respectively.}
\end{table}
Compared to the pair production threshold enforced by $R$-conserving SUSY models, the RPV mechanism would dramatically
reduce the energy scale of searching new physics. Driven by the $B$-violating $\hat{U}\hat{D}\hat{D}$ interactions, bottom
squark can be singly produced via quark-quark annihilation at the LHC. However, stringently
constrained by neutron-antineutron oscillation measurements, the $B$-violating coupling $\lambda_{113}^{\prime \prime}$ is too small
($| \lambda_{11k}^{\prime \prime} | < (10^{-8} - 10^{-7}) \frac{10^8~ {\rm s}}{\tau_{osc}} \left( \frac{\tilde{m}}{100~ {\rm GeV}} \right)^{5/2}$)
to be taken into account \cite{Zwirner}, while the other contributions from
$\lambda_{123}^{\prime \prime}$, $\lambda_{213}^{\prime \prime}$ and $\lambda_{223}^{\prime \prime}$ terms would be suppressed
by the smallness of sea quark parton densities. What's more, in the most simple case that the decay of the sbottom is driven
by the same RPV interaction as its production, the di-jet final state of the sbottom would be overwhelmed by QCD irreducible
background. Therefore, the direct search for the sbottom single production at the LHC might not be practicable,
especially in the case of only one $B$-violating coupling being nonzero.
Fortunately, additional to the current proton-proton collision mode, there are other beam configurations suggested to be possibly operated
in the same LHC accelerator. By colliding the existing $7~{\rm TeV}$ proton to an energetic new electron beam, the Large Hadron
Electron Collider (LHeC) \cite{LHeC-1,LHeC-2} is proposed to extend the deep inelastic lepton-hadron scattering into an unexplored
physics and kinematics, to improve the understanding of parton densities of proton at large four-momentum transfer squared and
low Bjorken-$x$. The electron beam of the LHeC is set to in the range of $50 - 150~ {\rm GeV}$ at an instant luminosity of about
$10^{33} ~ {\rm cm^{-2}s^{-1}}$. The capability of high energy and luminosity $e p$ collision of the LHeC makes it not only
feasible to study parton distributions of proton in an unexplored kinematic region, but also uniquely sensitive to the direct
single production of massive exotic electron-quark bound states \cite{weiht,kuday}. In this paper, we would discuss the potential
of probing the signal of the lighter sbottom resonance in the electron and jet final state at the LHeC, where both the production
and decay are governed by the $\hat{L}\hat{Q}\hat{D}$ $\lambda^{\prime}_{113}$-type Yukawa interaction. By comparing the cross sections for the sbottom pair production
at the LHC (${\rm Table~ \ref{table0-LOcs-LHC}}$) with those for the single sbottom production at the LHeC (${\rm Table~ \ref{table2-cs-lumi}}$),
we may conclude that the $E_e = 150~{\rm GeV}$ LHeC would provide excellent opportunities to search for the sbottom resonance and
probe the $L$-violating Yukawa coupling $\lambda^{\prime}_{113}$, especially in high sbottom mass region.

\section{RPV signal at the LHeC}
\par
The nonzero trilinear lepton-flavor-violating couplings $\lambda_{1j3}^{\prime}$ would induce resonant production and cascade decay
of sbottom at the LHeC via the $e^- p \rightarrow \tilde{b}_R \rightarrow e^- + jet + X$ process. Out of the two possible
$\hat{L}\hat{Q}\hat{D}$ couplings $\lambda^{\prime}_{1j3}~(j=1,2)$, the contribution via $\lambda^{\prime}_{123}$ is suppressed
by the smallness of charm quark parton density. Therefore, we restrict our discussion under the single coupling dominance hypothesis \cite{Barbier}
that all the RPV couplings except $\lambda_{113}^{\prime}$ are zero, and assume the lighter sbottom as the LSP. As a consequence,
only the valance up quark scattering will contribute to the LSP resonance, which can only decay into $e^- + u$ final state.

\par
\begin{figure}[htbp]
 \vspace*{-5.1cm}
 \centering
 \includegraphics[scale=1]{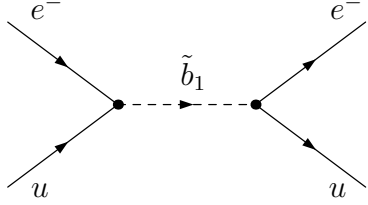}
 \vspace*{-22cm}
 \centering
 \caption{Feynman diagram for the parton level RPV signal process
 $e^- u \rightarrow \tilde{b}_1 \rightarrow e^- u$.}
 \label{Fig1-RPV-feynman}
\end{figure}

The parton level RPV signal process at the LHeC can be expressed as
\begin{eqnarray}
 e^-(p_1) + u(p_2) \rightarrow \tilde{b}_1 \rightarrow e^-(p_3) + u(p_4),
\end{eqnarray}
where $p_i~(i=1,2,3,4)$ are the four-momenta of initial and final particles, respectively. The Feynman diagram for this partonic
process is depicted in Fig.\ref{Fig1-RPV-feynman}.
The amplitude for the signal process at parton level has the form as
\begin{eqnarray}
 {\cal M}_{RPV} &=&
 -\vert \lambda^{\prime}_{113} \vert^2
 \sin^2\theta_{\tilde{b}}
 \left[
 \overline{u_e^c}(p_1) \frac{1 - \gamma^5}{2} u_u(p_2)
 \right]
 \frac{i}{\hat{s} - m_{\tilde{b}_1}^2 + i m_{\tilde{b}_1} \Gamma_{\tilde{b}_1}}
 \left[
 \overline{u}_u(p_4) \frac{1 + \gamma^5}{2} u_e^c(p_3)
 \right] \\
 &\stackrel{\mathrm{Fierz}}{\longrightarrow}&
 -\frac{\vert \lambda^{\prime}_{113} \vert^2}{2}
 \sin^2\theta_{\tilde{b}}
 \left[
 \overline{u}_e(p_3) \gamma^{\mu} \frac{1 - \gamma^5}{2} u_e(p_1)
 \right]
 \frac{i}{\hat{s} - m_{\tilde{b}_1}^2 + i m_{\tilde{b}_1} \Gamma_{\tilde{b}_1}}
 \left[
 \overline{u}_u(p_4) \gamma_{\mu} \frac{1 - \gamma^5}{2} u_u(p_2)
 \right], \nonumber
\end{eqnarray}
where $\sqrt{\hat{s}}$ is the center-of-mass (c.m.) colliding energy of the hard scattering and equivalent to the final state
invariant mass, and $\theta_{\tilde{b}}$ the sbottom mixing angle defined as
\begin{eqnarray}
 \left(
 \begin{array}{c}
 \tilde{b}_1 \\
 \tilde{b}_2
 \end{array}
 \right)
 =
 \left(
 \begin{array}{lc}
 ~\cos\theta_{\tilde{b}}~ & ~\sin\theta_{\tilde{b}}~ \\
 -\sin\theta_{\tilde{b}}~ & ~\cos\theta_{\tilde{b}}~
 \end{array}
 \right)
 \left(
 \begin{array}{c}
 \tilde{b}_L \\
 \tilde{b}_R
 \end{array}
 \right).
\end{eqnarray}
Then the differential cross section for the parton level signal process in the c.m. system can be expressed as
\begin{eqnarray}
 \frac{d \hat{\sigma}}{d \Omega}
 =
 \frac{1}{256 \pi^2} \vert \lambda^{\prime}_{113} \vert^4 \sin^4\theta_{\tilde{b}}
 \frac{\hat{s}}{(\hat{s} - m_{\tilde{b}_1}^2)^2 + m_{\tilde{b}_1}^2 \Gamma_{\tilde{b}_1}^2},
\end{eqnarray}
where the total decay width of the lighter sbottom, $\Gamma_{\tilde{b}_1}$, can be written out as
\begin{eqnarray}
 \Gamma_{\tilde{b}_1}
 =
 \frac{1}{16 \pi} \vert \lambda^{\prime}_{113} \vert^2 \sin^2\theta_{\tilde{b}} m_{\tilde{b}_1}.
\end{eqnarray}
In this paper, we take $\sin\theta_{\tilde{b}} = 1$ and therefore $\tilde{b}_1 = \tilde{b}_R$, by assuming that $m_b = 0$ and
$m_{\tilde{b}_R} < m_{\tilde{b}_L}$.

\par
For the parent level signal process $e^- p \rightarrow \tilde{b}_1 \rightarrow e^- + jet + X$, the kinematic distributions and
integrated cross section can be obtained by convoluting the parton level process with the parton distribution function (PDF) \cite{Placakyte}
of up quark in the proton,
\begin{eqnarray}
 d \sigma(e^- p \rightarrow \tilde{b}_1 \rightarrow e^- + jet + X) =
 \int dx G_{u/P}(x, \mu_f)
 d \hat{\sigma}(e^- u \rightarrow \tilde{b}_1 \rightarrow e^- u, \sqrt{\hat{s}} = 2\sqrt{x E_e E_p}).~~~~
\end{eqnarray}

\par
The RPV signal is dominated by the $s$-channel resonant production, and thus dramatically enhanced and sharply peaked around
the sbottom mass in the final state invariant mass spectrum in case of narrow decay width of the LSP. Another feature of the
RPV signal is that, attributed to the $L$-violating $\hat{L} \hat{Q} \hat{D}$ Yukawa interaction, the final state outgoing
particles are isotropic in the rest frame of sbottom resonance, which makes them energetic in the transverse momentum spectrum.

\par
The $e^- + jet$ background against signal mainly comes from the following eight SM partonic processes:
\begin{eqnarray}
 e^- + q \rightarrow e^- + q,~~~~~~ (q = u,\bar{u},d,\bar{d},c,\bar{c},s,\bar{s}).
\end{eqnarray}
Contrary to the signal $s$-channel production, all the SM background processes are governed by the $t$-channel electroweak
$Z/\gamma$ exchange. Consequently, the cross section for the SM deep inelastic scattering decreases quickly with the increment
of the invariant mass of final products, and both the outgoing electron and jet are more forward/backward than isotropic in
the c.m. frame of the final state. The distinctions between the RPV signal and the SM background could be useful in developing
an experimental event selection algorithm to improve the significance of sbottom resonance.

\section{Results and discussion}
\par
We generate Monte Carlo (MC) samples to simulate the RPV signal and the SM background, compare kinematic distributions,
and develop an optimized event selection algorithm to improve the sensitivity of searching for the lighter sbottom resonance
in $e^- + jet$ production at the proposed LHeC. For the RPV signal, there are only two free model input parameters, namely
the $\hat{L}\hat{Q}\hat{D}$ coupling constant $\lambda^{\prime}_{113}$ and the lighter sbottom mass $m_{\tilde{b}_1}$.
The most stringent $2 \sigma$ bound on the $L$-violating coupling $\lambda^{\prime}_{113}$, obtained from the CKM matrix
element $V_{ud}$ based on the single RPV coupling dominance hypothesis, is \cite{Barbier}
\begin{eqnarray}
 \label{upper limit from CKM matrix}
 \lambda^{\prime}_{113} < 0.02 \times \frac{m_{\tilde{b}_R}}{100~{\rm GeV}}.
\end{eqnarray}
In this paper, these two parameters are constrained in the range of
\begin{eqnarray}
\lambda^{\prime}_{113} \leq 0.02,~~~~~ m_{\tilde{b}_1} \geq 100~{\rm GeV}.
\end{eqnarray}
This set of parameters will be used as default unless stated otherwise. To generate background events, SM input parameters are
taken as $m_u = m_d = 0$, $m_c = 1.275~ {\rm GeV}$, $m_s = 95~{\rm MeV}$, $m_W = 80.385~ {\rm GeV}$, $m_Z = 91.1876~ {\rm GeV}$,
$\Gamma_Z = 2.4952~ {\rm GeV}$ and $\alpha_{{\rm ew}} = 1/127.944$ \cite{Beringer}. {\it FeynArts-3.4} package \cite{FeynArts}
is used to generate the Feynman diagrams for the RPV signal and the SM background processes at parton level and convert them
into the corresponding amplitudes, and {\it FormCalc-5.4} package \cite{FormCalc} is subsequently employed to simplify the
Feynman amplitudes and generate the MC events. For the initial parton convolution, we adopt the {\it CTEQ6L1} PDFs \cite{CTEQ}
and set the factorization scale $\mu_F$ as the lighter sbottom mass $m_{\tilde{b}_1}$ and the final state invariant mass $M_{\mbox{e-jet}}$
for the RPV signal and the SM background, respectively. When generating event samples and calculating cross sections,
a set of kinematic and geometric acceptance requirements on transverse momentum $p_T$ and rapidity $y$ in the lab frame,
\begin{eqnarray}
 \label{kinematic_cuts-I}
 p_{T}^{e} > 20~{\rm GeV}, ~~~~ p_{T}^{jet} > 20~{\rm GeV},~~~~ |y_{e}| < 3.2, ~~~~ |y_{jet}| < 4.9,
\end{eqnarray}
is imposed as baseline cuts, where the $z$-axis is defined in direction of the electron beam.

\par
The cross sections for the RPV signal with the baseline requirements of Eq.(\ref{kinematic_cuts-I}) are compared with the SM
predictions in ${\rm Table~ \ref{table-cs-baseline}}$, with two proposed LHeC electron beam energy configurations of
$E_e = 50~ {\rm GeV}$ and $E_e = 150~ {\rm GeV}$, respectively. With the basic selection, the production rates for the SM
background are about $3 - 4$ orders of magnitude larger than what are expected from the RPV signal. However, driven by
different mechanics of interactions, there are distinctions between the RPV signal and the SM background.
\begin{table}[htbp]
\centering
\begin{tabular}{|c|c|c|c|}
 \hline
 \multirow{2}*{$E_e~ ({\rm GeV})$} & \multicolumn{3}{|c|}{$\sigma~ (pb)$} \\
 \cline{2-4}
 & $m_{\tilde{b}_1} = 150~ {\rm GeV}$ & $m_{\tilde{b}_1} = 500~ {\rm GeV}$ & SM background \\
 \hline
 50 & 3.49 & 0.214 & 2853.6  \\
 \hline
 150 & 4.69 & 0.278 & 5013.6  \\
 \hline
\end{tabular}
 \caption{
 \label{table-cs-baseline}
 Cross sections for RPV signal and SM background with baseline cuts at the LHeC with $50~{\rm GeV}$ and $150~{\rm GeV}$ electron beams.}
\end{table}

\par
\begin{figure}[htbp]
 \centering
 \includegraphics[scale=0.50]{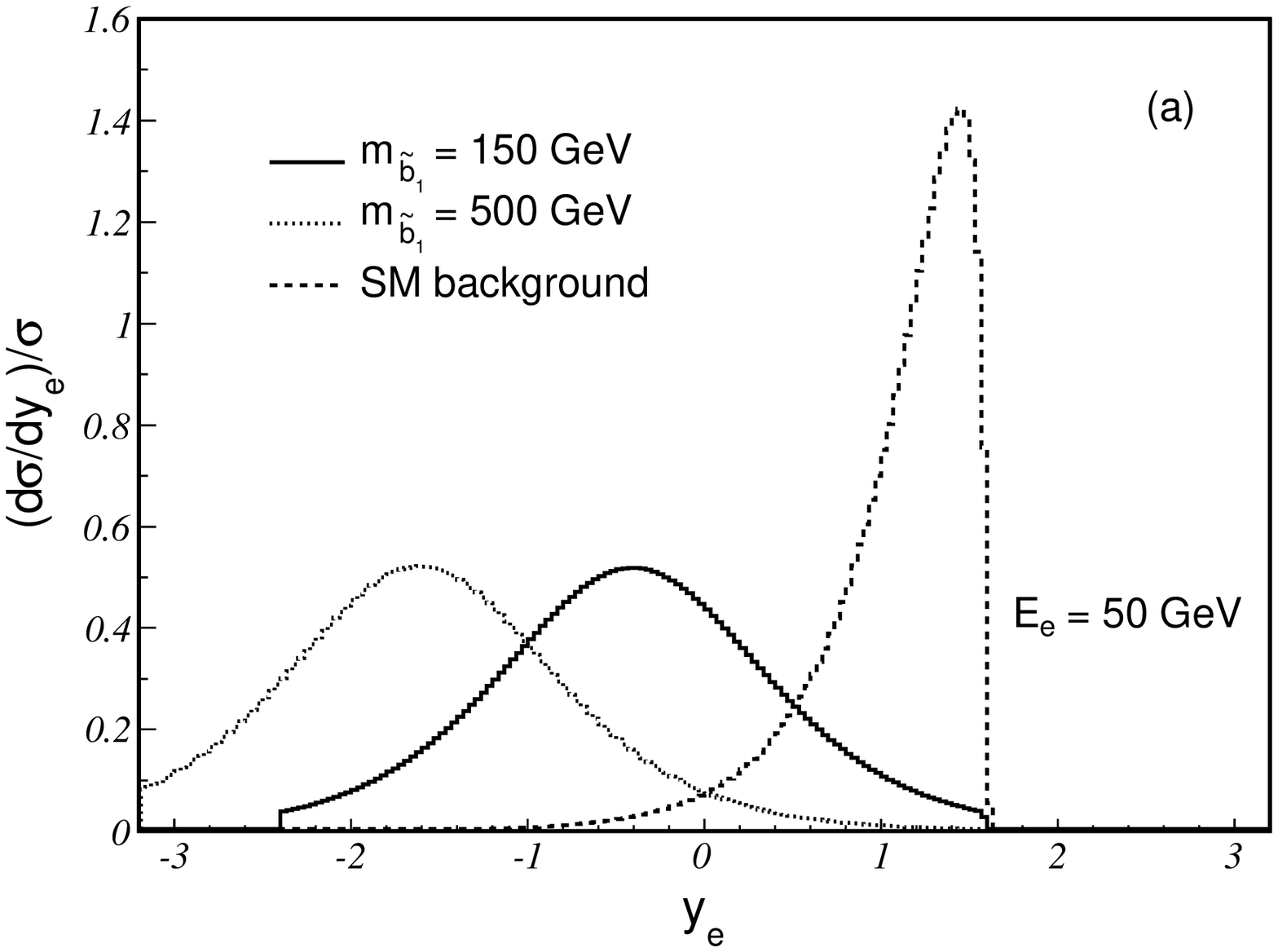}
 \includegraphics[scale=0.50]{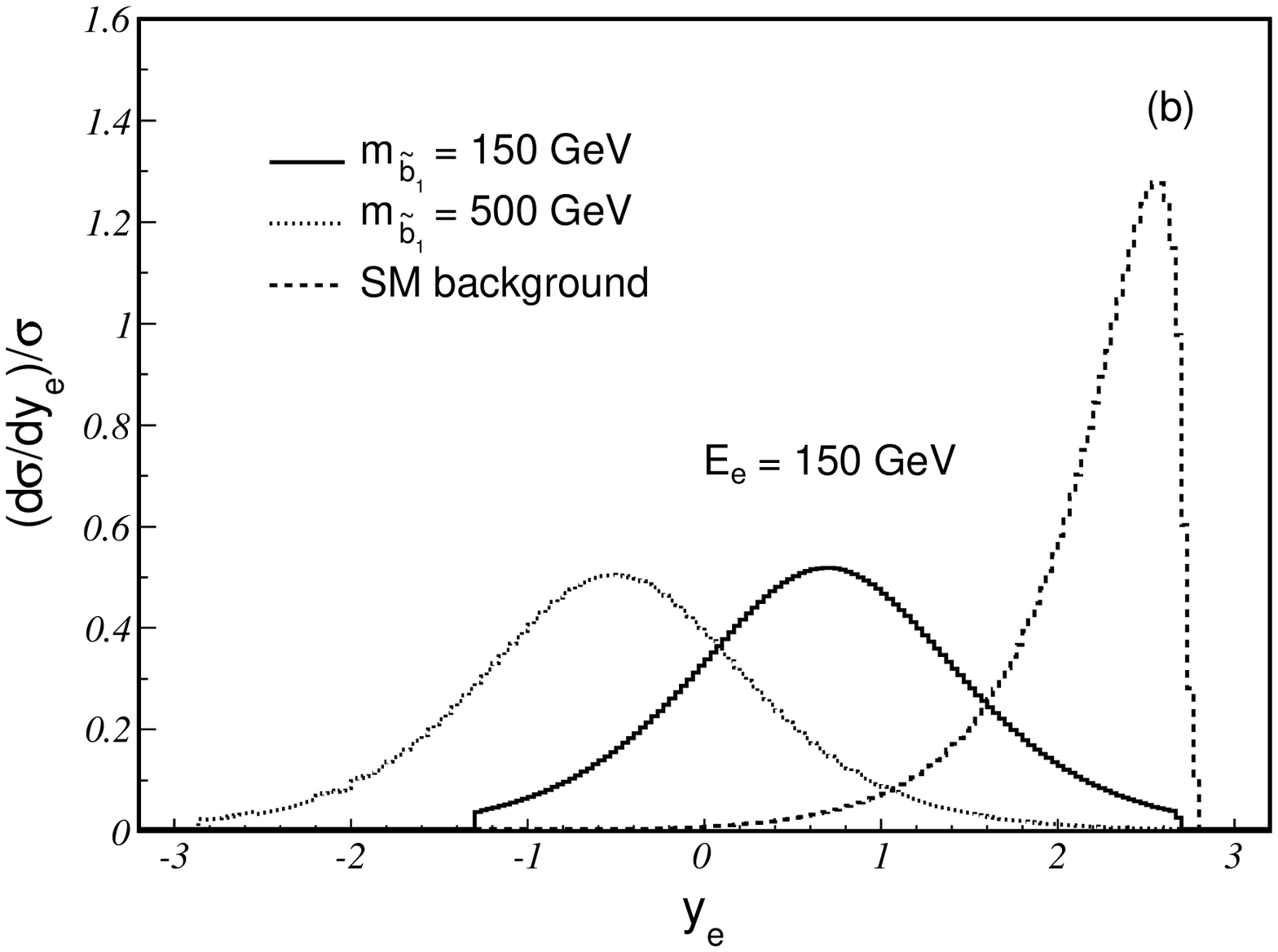}
 \centering
 \caption{Normalized distributions of the final state electron rapidity in the lab frame
 with baseline cuts at the LHeC with (a) $50~ {\rm GeV}$ and (b) $150~ {\rm GeV}$ electron beams.}
 \label{Fig2-Ey-lab}
\end{figure}
\begin{figure}[htbp]
 \centering
 \includegraphics[scale=0.50]{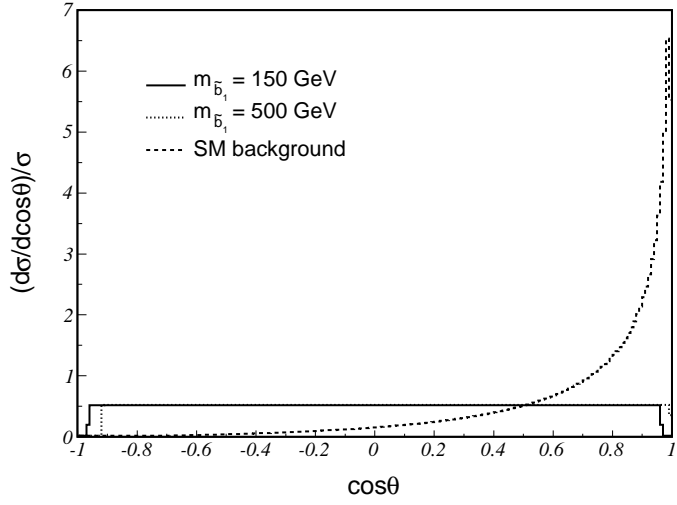}
 \centering
 \caption{Normalized distributions of the electron scattering angle in the c.m. frame of
 the final state with baseline cuts at the LHeC with $50~ {\rm GeV}$ electron beam.}
\label{Fig3-Etheta-cms-Ee50}
\end{figure}

First, the scalar nature of RPV Yukawa couplings makes the outgoing electron and jet of the $s$-channel RPV signal process isotropic,
while the final particles of the SM deep inelastic scattering tend to be forward/backward with respect to the beam direction, in the
c.m. frame of the final state. This trend diminishes in the lab frame due to the heavy boost along the incoming proton or electron
beam direction, depending on the electron beam energy and the mass of sbottom resonance, as shown in Fig.\ref{Fig2-Ey-lab}.
The rapidity distributions for the RPV signal in the lab frame vary with the lighter sbottom mass and the electron beam configuration,
and would be mixed with those for the SM background.
On the other hand, transferred back to the c.m. frame of the final state, the distributions of the electron scattering angle for the
RPV signal are independent of the colliding energy and the $\slashed{R}_{p}$-MSSM input parameters, and are distinguishable from the
SM deep inelastic scattering, as depicted in Fig.\ref{Fig3-Etheta-cms-Ee50}. Therefore, a uniform cut on the electron scattering angle
in the c.m. frame of the final state would efficiently separate the RPV signal from the SM background.

\par
\begin{figure}[htbp]
 \centering
 \includegraphics[scale=0.50]{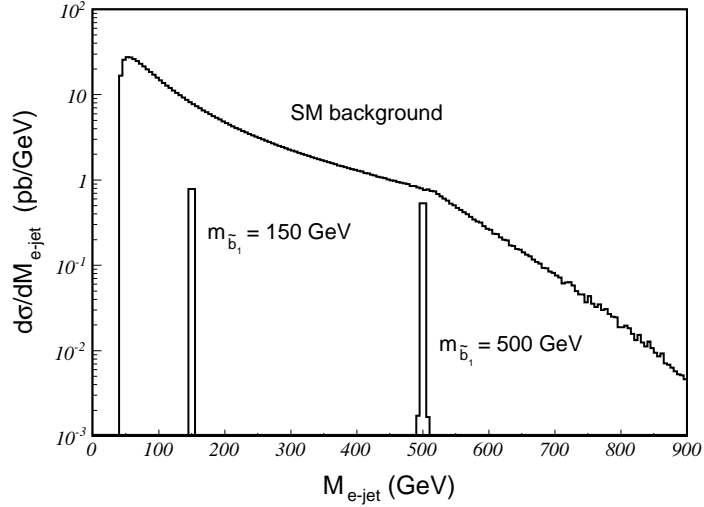}
 \centering
 \caption{Distributions of the final state invariant mass with baseline cuts at the LHeC
 with $50~ {\rm GeV}$ electron beam.}
\label{Fig4-Mej-Ee50}
\end{figure}
\begin{figure}[htbp]
 \centering
 \includegraphics[scale=0.50]{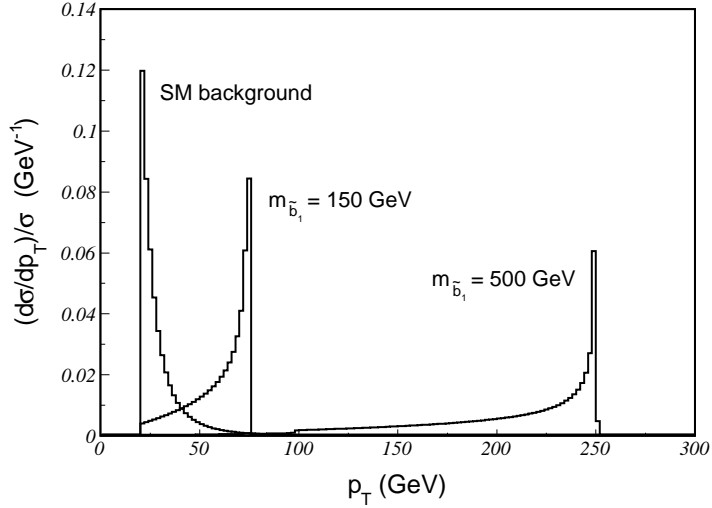}
 \centering
 \caption{Normalized distributions of the electron/jet transverse momentum with baseline
 cuts at the LHeC with $50~ {\rm GeV}$ electron beam.}
\label{Fig5-epT-Ee50}
\end{figure}

Second, due to the weak $\hat{L}\hat{Q}\hat{D}$ coupling strength, the total decay width of the lighter sbottom is only about ${\cal O}(1)~ {\rm MeV}$,
and consequently will not only enhance the signal production rate via resonance effect, but also signify itself as a striking peak in
the final state invariant mass spectrum. This narrow resonance effect, together with the high transverse momentum tendency of the RPV
signal against the SM background, are depicted in Fig.\ref{Fig4-Mej-Ee50} and Fig.\ref{Fig5-epT-Ee50}, respectively.

\par
\begin{figure}[ht]
 \centering
 \includegraphics[scale=0.50]{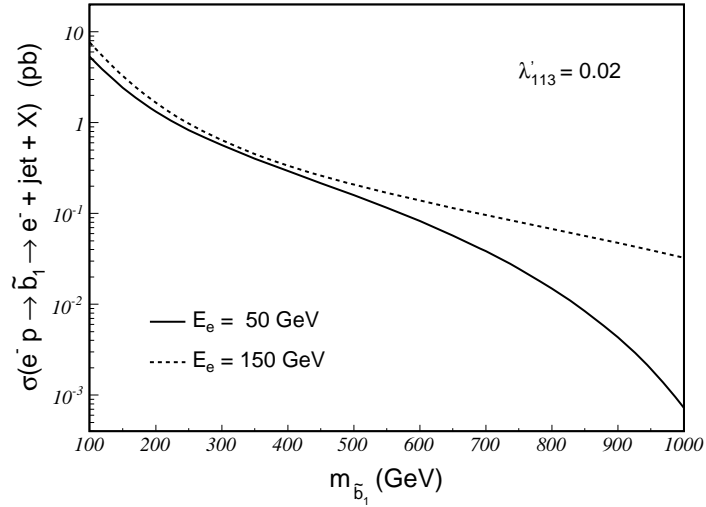}
 \centering
 \caption{Cross sections for the lighter sbottom resonance production at the LHeC after the event selection criteria have been applied.}
\label{Fig6-sigx-allcuts}
\end{figure}

Based on the above consideration, an optimized event selection criteria could be developed by tightening transverse momentum cuts,
separating the electron scattering angle of the signal from that of the background in the c.m. frame of the final state, and imposing
a stringent invariant mass window around the sbottom signal, and is finalized as follows.
\begin{itemize}
\item Rapidity cuts: $|y_{e}| < 3.2$, $|y_{jet}| < 4.9$;

\item Transverse momentum cuts: $p_{T}^{e} > 40~{\rm GeV}$, $p_{T}^{jet} > 40~{\rm GeV}$;

\item Electron scattering angle cut: $\cos\theta < 0.5$;

\item Invariant mass cut: $|M_{\mbox{e-jet}} - m_{\tilde{b}_1}| < 5\% m_{\tilde{b}_1}$.
\end{itemize}
In this event selection criteria, the rapidity and electron scattering angle are defined in the lab frame and the c.m. frame of the
final state, respectively. The lighter sbottom decay width is negligible compared with typical experimental resolution, which is
quantified as $5\%$ dominated by $30\%/\sqrt{E_{jet}}$ hadronic calorimeter resolution in unit of GeV. And the isotropy of final
products of the RPV signal in the c.m. frame of the final state makes $p_T^{e/jet}$ tend to take the half of the sbottom energy as
about $0.5 m_{\tilde{b}_1}$. Accordingly, signal events will pass the invariant mass and transverse momentum cuts with almost $100\%$
efficiency. Only one fourth signal events will be lost through the electron scattering angle cut. The RPV signal cross sections
after all the above event selection cuts have been applied are depicted in Fig.\ref{Fig6-sigx-allcuts}, for both $E_e = 50~ {\rm GeV}$
and $E_e = 150~ {\rm GeV}$ electron beam options. On the other hand, the cross sections for the SM deep inelastic scattering after
the event selection criteria have been applied could be heavily suppressed by a factor of ${\cal O}(10^3 - 10^4)$, and are thus
comparable to the RPV signal.

\par
\begin{figure}[ht]
 \centering
 \includegraphics[scale=0.50]{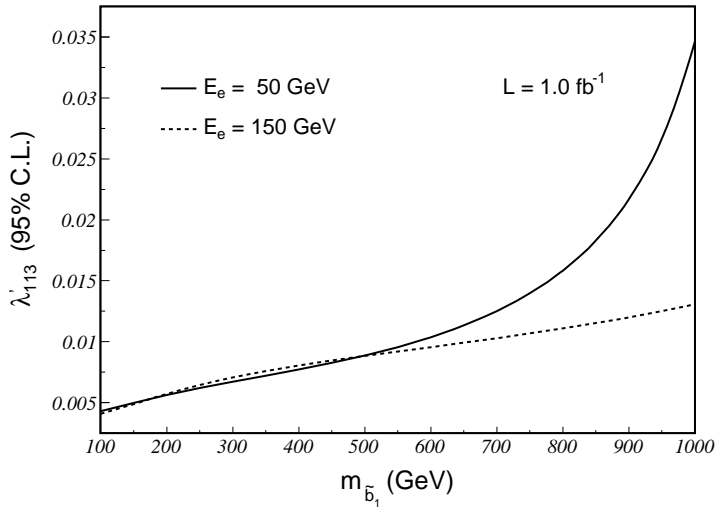}
 \centering
 \caption{$95\%$ C.L. upper bounds on $\lambda^{\prime}_{113}$ as functions of $m_{\tilde{b}_1}$.}
 \label{Fig7-bounds}
\end{figure}

The significance of signal over background $S$ is defined as
\begin{eqnarray}
S = \frac{N_S}{\sqrt{N_B}} = \frac{\sigma_S}{\sqrt{\sigma_B}} \cdot \sqrt{\cal L},
\end{eqnarray}
where $N_{S,B}$ and $\sigma_{S,B}$ are the numbers of events and the cross sections for signal and background, and ${\cal L}$ denotes
the accumulated luminosity.
\begin{table}[htb]
  \centering
  \begin{tabular}{c|ccccc}
  \hline
  \multirow{2}*{$m_{\tilde{b}_1}~({\rm GeV})$} &
  \multicolumn{2}{c}{$E_e = 50~{\rm GeV}$}  &&
  \multicolumn{2}{c}{$E_e = 150~{\rm GeV}$}  \\
  &  $\sigma_{S}~(pb)$
  &  ${\cal L}~(pb^{-1})$
  &  ~~~~~~
  &  $\sigma_{S}~(pb)$
  &  ${\cal L}~(pb^{-1})$  \\
  \hline
  100  &  $5.34$                 &  $13.5$                &&  $7.69$                 &  $10.8$  \\
  200  &  $1.32$                 &  $39.7$                &&  $1.66$                 &  $42.0$  \\
  300  &  $0.566$                &  $79.2$                &&  $0.641$                &  $97.6$  \\
  400  &  $0.293$                &  $139$                 &&  $0.337$                &  $164$  \\
  500  &  $0.159$                &  $241$                 &&  $0.208$                &  $238$  \\
  600  &  $8.24 \times 10^{-2}$  &  $451$                 &&  $0.139$                &  $326$  \\
  700  &  $3.83 \times 10^{-2}$  &  $956$                 &&  $9.61 \times 10^{-2}$  &  $438$  \\
  800  &  $1.49 \times 10^{-2}$  &  $2.46 \times 10^{3}$  &&  $6.75 \times 10^{-2}$  &  $591$  \\
  900  &  $4.31 \times 10^{-3}$  &  $8.64 \times 10^{3}$  &&  $4.73 \times 10^{-2}$  &  $810$  \\
  1000 &  ~$7.23 \times 10^{-4}$~  &  $5.62 \times 10^{4}$  &&  ~$3.25 \times 10^{-2}$~  &  $1.14 \times 10^{3}$  \\
  \hline
  \end{tabular}
  \caption{\label{table2-cs-lumi}
  Cross sections for the RPV signal after the event selection criteria have been applied and
  luminosities required for the $S \geq 5$ discovery of the lighter sbottom resonance at the
  LHeC with the electron beam energies of $E_e = 50~{\rm GeV}$ and $E_e = 150~{\rm GeV}$.}
\end{table}
${\rm Table~ \ref{table2-cs-lumi}}$ gives the cross sections for the RPV signal after the event selection criteria have been applied and the
luminosities needed to declare the $S \geq 5$ discovery of the lighter sbottom resonance via $e^- + jet$ final state at the LHeC,
with $\lambda^{\prime}_{113} = 0.02$, for both the low and the high electron beam energy configurations. It can be seen that the
$E_e=150~{\rm GeV}$ electron beam option is more powerful than the $E_e=50~{\rm GeV}$ configuration in searching for the lighter
sbottom resonance RPV signal, particularly in the mass range of $m_{\tilde{b}_1} > 700~{\rm GeV}$. If there is no significant
deviation from the SM prediction on the deep inelastic scattering observed with certain accumulated luminosity at the LHeC,
e.g. ${\cal L} = 1~fb^{-1}$, the exclusion on the $L$-violating coupling $\lambda_{113}^{\prime}$ can be drawn at $95\%$ confidence
level (C.L.), as shown in Fig.\ref{Fig7-bounds}.
It shows that the direct search for the lighter sbottom resonance via $e^- + jet$ final state at the LHeC would give much more stringent
limit on the $L$-violating coupling $\lambda^{\prime}_{113}$ than the low energy experiments, especially in high sbottom mass region.
For example, at $m_{\tilde{b}_1}=1~{\rm TeV}$ mass point, a new $95\%$ C.L. upper bound on $\lambda^{'}_{113}$ can be derived as small
as $0.013$, at the LHeC with $1~fb^{-1}$ data expected in $150~{\rm GeV}$ electron beam collision, which is about 15 times smaller than
the $2 \sigma$ upper limit obtained from the $V_{ud}$ measurement given by Eq.(\ref{upper limit from CKM matrix}).

\vskip 5mm
\par
Some issues should be addressed here:
\begin{itemize}
  \item
  The resonant productions of top and bottom squarks at the LHeC have been investigated in Refs.\cite{weiht,kuday} via $\mu + jet$ final state.
  Due to the lepton-flavor violation, two different $L$-violating $\hat{L}\hat{Q}\hat{D}$ Yukawa couplings (with different lepton flavors) are involved
  in the signal processes and there is no SM irreducible background. Only the upper bounds on the quadratic $L$-violating coupling products can be obtained
  if those squark resonances are not discovered. In our paper, we study the sbottom resonance production at the LHeC via a lepton-flavor-conserving process
  under the single coupling dominance hypothesis.
  Although the SM irreducible background is huge, the signal events can be selected by taking appropriate cuts on final particles.
  Different from Refs.\cite{weiht,kuday}, an upper bound on single RPV coupling can be obtained if there is no significant deviation from the SM prediction.
  \item
  In this paper, we do not consider the contributions of the $t$-channel RPV Feynman diagram and the interference between the RPV and
  the SM amplitudes, because these contributions are negligible after the event selection criteria have been applied.
  \begin{table}[htb]
  \centering
  \begin{tabular}{c|ccccc}
  \hline
  \multirow{2}*{$m_{\tilde{b}_1}~({\rm GeV})$} &
  \multicolumn{2}{c}{$E_e = 50~{\rm GeV}$}  &&
  \multicolumn{2}{c}{$E_e = 150~{\rm GeV}$}  \\
  &  $\delta \sigma_{S}/\sigma_{S}$
  &  $\delta {\cal L}/{\cal L}$
  &  ~~~~~~
  &  $\delta \sigma_{S}/\sigma_{S}$
  &  $\delta {\cal L}/{\cal L}$  \\
  \hline
  100  &  $6.80 \times 10^{-4}$    &  $-1.36 \times 10^{-3}$   &&  $7.50 \times 10^{-4}$    &  $-1.50 \times 10^{-3}$  \\
  400  &  $1.05 \times 10^{-3}$    &  $-2.11 \times 10^{-3}$   &&  $9.66 \times 10^{-4}$    &  $-1.93 \times 10^{-3}$  \\
  700  &  $3.09 \times 10^{-3}$    &  $-6.15 \times 10^{-3}$   &&  $1.12 \times 10^{-3}$    &  $-2.24 \times 10^{-3}$  \\
  1000 &  ~$1.36 \times 10^{-2}$~  &  $-2.67 \times 10^{-2}$   &&  ~$1.91 \times 10^{-3}$~  &  $-3.81 \times 10^{-3}$  \\
  \hline
  \end{tabular}
  \caption{\label{table3-correction-cs-lumi}
  Relative corrections induced by the contributions of the $t$-channel RPV Feynman diagram
  and the interference between the RPV and the SM amplitudes to the RPV signal cross section
  and the luminosity required for the $S \geq 5$ discovery of the lighter sbottom resonance
  at the LHeC with the electron beam energies of $E_e = 50~{\rm GeV}$ and $E_e = 150~{\rm GeV}$.}
  \end{table}
  As shown in ${\rm Table~ \ref{table3-correction-cs-lumi}}$, the relative corrections induced by these contributions to the RPV signal
  cross section and the luminosity required for the $S \geq 5$ discovery of the lighter sbottom resonance, $\delta \sigma_S/\sigma_S$
  and $\delta {\cal L}/{\cal L}$, are almost less then $1\%$. Therefore, the numerical results and conclusion of the paper are reasonable.
  \item
  The above lighter sbottom resonance searching strategy can be easily translated to probing the lighter scalar top $\tilde{t}_1$
  (assuming to be the LSP) and the lepton-number-violating $\lambda^{\prime}_{131}$-type Yukawa interaction,
  via $e^+ + d \rightarrow \tilde{t}_1 \rightarrow e^+ + d$ process in $e^+ p$ collision at the LHeC. However, due to the lower luminosity
  of positron beam and the smaller fraction of energetic valence $d$-quark density in proton against electron beam and $u$-quark density,
  the potential of the discovery of the lighter stop resonance or the constraint on the relevant RPV coupling in $e^+ p$ scattering would
  be diminished, compared to the sbottom resonance search in $e^-p$ collision mode.
\end{itemize}

\section{Summary}
\par
In this paper, the possibility of probing the lepton-number-violating $\hat{L}\hat{Q}\hat{D}$ $\lambda^{\prime}_{113}$-type interaction
in the deep inelastic scattering at the LHeC is investigated. Under the single coupling dominance hypothesis that all the RPV couplings
except $\lambda^{\prime}_{113}$ are zero, the lighter sbottom can be produced and subsequently decay into electron and jet. Taking advantage
of the enhancement of the narrow width resonant production of sbottom and distinctive kinematic distributions of $\tilde{b} \rightarrow e^- + u$
decay products in the c.m. frame of the two-body final state, an optimized event selection strategy can be developed to improve the
sensitivity of the RPV signal over SM background. With about $1~fb^{-1}$ integrated luminosity expected in $E_e = 150~{\rm GeV}$ $e^-p$
collision at the LHeC, either the lighter sbottom RPV resonance could be directly detected up to $1~{\rm TeV}$, or a new $95\%$ C.L. upper
bound on the $L$-violating coupling $\lambda^{\prime}_{113}$, which is one order of magnitude smaller than that obtained from the CKM matrix
element $V_{ud}$, could be derived.

\vskip 5mm
\par
\noindent{\large\bf Acknowledgments:} This work was supported in part by the National Natural Science Foundation of China
(Grants No. 11075150, No. 11025528 and No. 11275190).

\end{document}